\documentclass[twocolumn,pra,aps,showpacs]{revtex4}
\usepackage{amsmath,amsfonts,amsthm,amssymb,amsbsy,graphicx,color}
\pdfoutput=1

\begin{document}

\def\ra{\rangle}
\def\la{\langle}
\def\vp{\varphi}
\def\be{\begin{equation}}
\def\ee{\end{equation}}
\def\ba{\begin{eqnarray}}
\def\ea{\end{eqnarray}}
\def\ap{\alpha}
\def\bt{\beta }
\newcommand{\bra}[1]{\left\langle #1 \right\vert}
\newcommand{\ket}[1]{\left\vert #1 \right\rangle}
\newcommand{\re}[1]{\langle #1 \rangle}
\newcommand{\bx}{\begin{matrix}}
\newcommand{\ex}{\end{matrix}}
\newcommand{\mc}[1]{\mathcal}

\title{Canonical Phase Measurements in the Presence of Photon Loss}
\author{Aravind Chiruvelli}
%\email{chiruvelli@phys.lsu.edu}% and Hwang Lee} 
\author{ Hwang Lee}
%\email{hwlee@phys.lsu.edu}

\affiliation{Hearne Institute for Theoretical Physics, Department of Physics and Astronomy\\Louisiana State University, Baton Rouge, LA 70803, USA}

%\author{Hwang Lee}
\date{\today}
\begin{abstract}
We analyze the optimal state under the canonical phase measurement, as given by Berry and Wiseman [Phys. Rev. Lett {\bf 85}, 5098, (2000)], in the presence of photon loss. The model of photon loss is a generic  fictitious beam splitter, and we present the full density matrix calculations, which are more direct and do not involve any approximations. We find for a given amount of loss the upper bound for the input photon number that yields a sub-shot noise estimate.
\end{abstract}


\maketitle

\section{Introduction}

Canonical phase measurement in quantum mechanics is a significant problem, for the main reason that phase is a quantity that is conjugate to the number, $N$, of photons in a particular electromagnetic mode~\cite{Lane,hradil-1,hradil-2}. Due to this conjugate nature, the phase estimate $\Delta\vp$ is ultimately limited by the number $N$ of the photons as $\Delta\vp = 1/N$, which is conventionally  referred as the Heisenberg limit. In the usual classical setting, such as interferometry with lasers, for a given number of input resources, $N$, phase estimate scales as $1/\sqrt{N}$, which is usually referred as shot-noise limit.

\begin{figure}[ht]
\includegraphics[height=0.75\hsize,width=1.0\hsize]{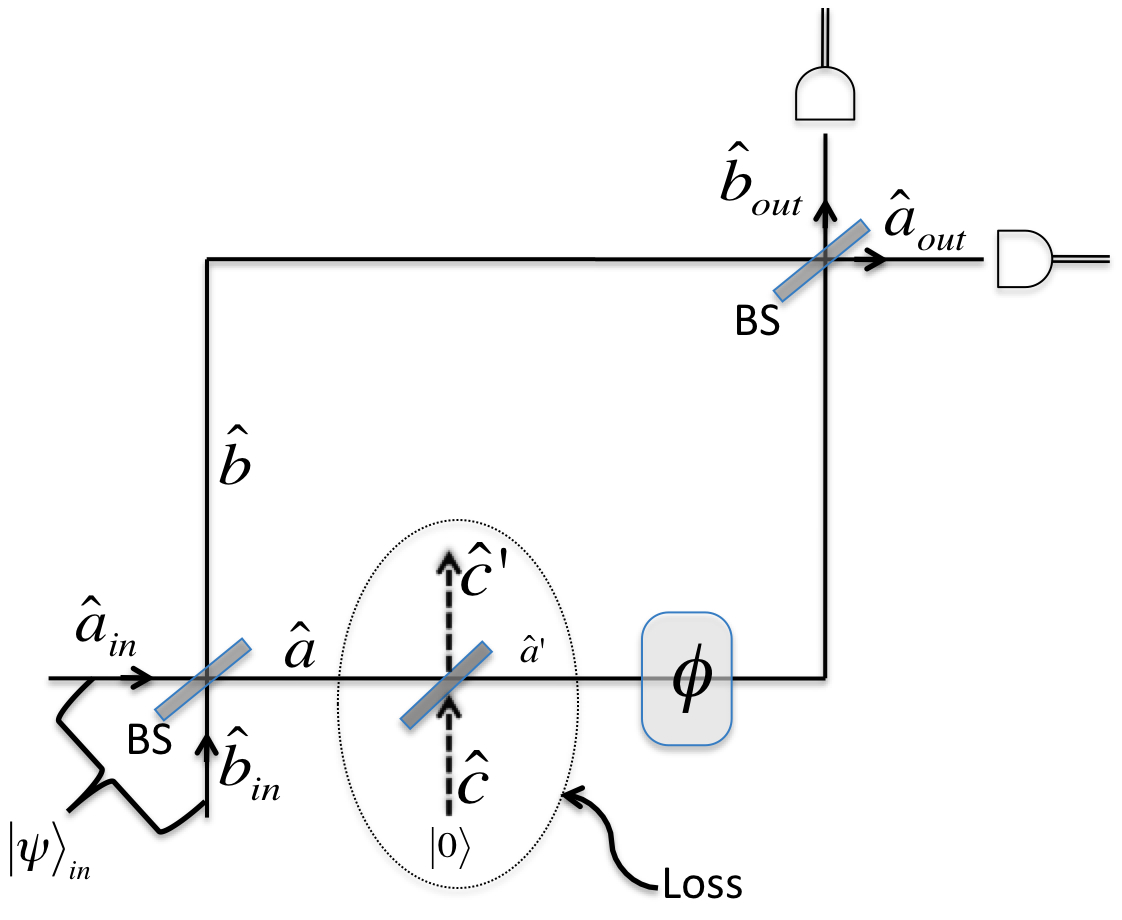}
\caption{Schematic of Mach-Zehender Interferometer (MZI) with photon loss at the phase shift. The state $|\psi\rangle_{\text{in}}$ represents the joint input state at $\hat{a}_{\text{in}}$ and $\hat{b}_{\text{in}}$. Photons in the lower arm first encounter the fictitious beam splitter (BS), for which vacuum enters through the other input, and depending the transmission coefficient, some of them are scattered into the mode $\hat{c}'$, which are ignored (traced out), and the remaining pass through the phase shifter.}
\label{mzi}
\end{figure}

 Accurate phase estimation has many practical applications such as metrology, imaging and sensing~\cite{seth}. Achieving Heisenberg limit in practice is not a trivial problem and there have been numerous proposals to achieve this limit~\cite{rosetta,noh,berry,cdual,Caves,cnoon,yuen,yurke,Bollinger:PR,jon98,Holl}.  Canonical phase measurement has been first dealt with Helstrom~\cite{helstrom} and Shapiro~\cite{shapiro}, and later Sanders and Milburn~\cite{opovm,sand-jmo} used it to obtain a phase estimate in a Mach-Zehender interferometer (MZI) as shown in Fig.\ref{mzi} (excluding the loss part). 
The phase estimate thus obtained is independent of the system phase, unlike in other methods~\cite{rosetta,noh,cdual} where the ultimate limit is achieved for a particular system phase. Also, the measurement specified by Sanders and Milburn is not particular to a specific input state. 

Motivated by this work, Berry and Wiseman~\cite{berry} analytically derived an input state, called as the optimal state, subject to the canonical measurement. They also  suggested an adaptive method of approximately implementing the canonical phase measurement~\cite{berry,berry-pra}, which has been  used in the recent experimental realization of the Heisenberg limited phase measurement by Higgins {\it et al.}~\cite{nature}.
The canonical measurement, written as  a Positive Operator Valued Measure(POVM) by Sanders and Milburn~\cite{opovm} is,
\be
\hat{F}(\vp)d\vp=\frac{2j+1}{2\pi}\ket{j\vp}\bra{j\vp}d\vp,
\label{opo}
\ee
in terms of the phase states 
\be
\ket{j\vp}=\frac{1}{\sqrt{2j+1}}\sum_{\ap=-j}^je^{i\ap\vp}\ket{j\ap}_y. \nonumber
\ee 

In defining this, the Schwinger's representation is used, and for completeness we wish to outline the notation. The three angular momentum components $\hat{J}_x, \hat{J}_y$ and $\hat{J}_z$ are very effective in analyzing two-port, lossless, interferometers~\cite{rosetta,opovm,noh}. For the two modes, $\hat{a}$ and $\hat{b}$ of the MZI (Fig.~\ref{mzi}), these two mode operators are,
\ba
\hat{J}_x=(\hat{a}^\dagger\hat{b}+\hat{a}\hat{b}^\dagger)/2, & & \hat{J}_y=(\hat{a}^\dagger\hat{b}-\hat{a}\hat{b}^\dagger)/2i,\nonumber \\ 
\hat{J}_z=(\hat{a}^\dagger\hat{a}-\hat{b}^\dagger\hat{b})/2, & & \hat{J}^2=\hat{J}_{x}^2+\hat{J}_{y}^2+\hat{J}_{z}^2. \nonumber
\ea

In the context of the MZI, $\hat{J}_x$ implements the operation of a 50-50 beam splitter as $e^{i(\pi/2)\hat{J}_x}$ and $\hat{J}_z$ defines the photon number difference in two modes. The simultaneous eigenvector of $\hat{J}^2$ and $\hat{J}_z$, $\ket{j,m}_z$ represents the joint input state $\ket{j+m}_{a_{in}}$ and $\ket{j-m}_{b_{in}}$ in the Fock-state basis, and the total input number of photons is $N=2j$. The simultaneous eigenvector of $\hat{J}^2$ and $\hat{J}_y$, which is $\ket{j,n}_y$, represents the joint state {\it within} the interferometer. The beamsplitter transformation in this representation performs a rotation about the $\hat{J}_x$. The phase states discussed above, are defined in terms of  states {\it within} the interferometer in Fig.~\ref{mzi} and thus the output modes or detectors are irrelevant for the present purpose. Thus the probability distribution for the system phase $\phi$ is obtained as
 \be
 P(\vp)d\vp=\bra{\psi}\hat{F}(\vp)\ket{\psi}d\vp=\text{Tr}[\rho\hat{F}(\vp)]d\vp.
 \label{prob}
 \ee
Note that $\vp$ is the estimate of the system phase $\phi$. The optimal state, to be specified below, is derived conditioned upon minimizing the Holevo variance calculated from the above probability distribution. The Holevo phase variance is defined as~\cite{holevo}
\be
(\Delta\vp)^2\equiv -1+|\la e^{i\vp} \ra|^{-2},
\label{varia}
\ee
where $|\la e^{i\vp} \ra|=\int_0^{2\pi}d\vp P(\vp)e^{i\vp-\bar{\vp}}$, is also called the sharpness. Here $\bar{\vp}$ is a mean phase and we take it to be zero. The Holevo variance for the optimal state is given by Ref.~\cite{berry,berry-pra},
\be
(\Delta\vp)^2=\tan^2(\frac{\pi}{N+2})\approx\frac{\pi^2}{N^2},
\label{hvar}
\ee
thus giving rise to a phase estimate that scales as the Heisenberg limit. Besides the optimal state, other prominent  states that achieve the Heisenberg limit are so called the NOON states~\cite{rosetta,Bollinger:PR} and the dual-fock state~\cite{cdual,cnoon}. Formally the NOON state is,
\be
\ket{\psi}_N =\frac{\ket{N}_{\hat{a}}\ket{0}_{\hat{b}}+\ket{0}_{\hat{a}}\ket{N}_{\hat{b}}}{\sqrt{2}}.
\ee
Note that the above state is not an input state at the MZI shown in Fig.1, but a state {\it within} the interferometer and the subscripts $\hat{a}$ and $\hat{b}$ denotes the internal modes of the MZI. The dual-fock input state is given as,
\be
\ket{\psi}_D = \ket{N}_{\hat{a}_{in}}\ket{N}_{\hat{b}_{in}}
\ee 

The detection scheme for both, the NOON and the dual-fock state that results in the Heisenberg limited phase estimate is the parity detection~\cite{cnoon,yang}, first proposed by Bollinger {\it et al.}~\cite{Bollinger:PR} in the context of frequency metrology with trapped ions. The shot-noise and sub shot-noise limit with matter wave interferometry which deals with Bose-Einstein condensates at the input of the MZI is studied in Refs.~\cite{pezze}. 

It is natural to question the performance of such states or the detection schemes in a more realistic conditions such as photon loss associated with the propagation. The analysis of the NOON states under propagation loss was carried out independently by Gilbert {\it et al.}~\cite{gilbert} and by Rubin and Kaushik~\cite{mark}, where they used pure state formalism. In Ref.~\cite{gilbert,mark} for NOON state, the minimum number of photons required to achieve a minimum detectable phase in presence of loss is also given.

In this paper we study the performance of optimal state and the optimal POVM in the presence of the  photon loss associated with propagation. We use the generic beam splitter model for photon loss~\cite{loudon}, as shown in Fig.~\ref{mzi}. The input mode $\hat{c}$ for this fictitious beam splitter is a vacuum mode, and the output mode $\hat{c}'$ is then to be traced out. This typically implies that the photons that are lost in mode $\hat{c}'$, due to the nonzero reflection coefficient $r$, correspond to the photon loss. 

In the following two Sections we describe optimal state in presence of photon loss and carry out the explicit density matrix calculation. In Section~\ref{conc}, we quantitatively describe the effect of photon loss on the canonical phase estimation. Section~\ref{conc1} concludes with numerical results and  discussions.

\section{Optimal state in presence of photon loss}

 We now proceed to develop the mathematical framework to study optimal-state canonical  interferometry  with photon loss. The beam splitter representing loss, with arbitrary transmission and reflection, can be characterized by an angle $\theta$, such that transmission and reflection coefficients are $\tau=\cos^2(\theta/2)$ and $r=1-\cos^2(\theta/2)$, respectively. Therefore, the loss is simply the reflection coefficient, 
 \be
 L=r=1-\cos^2(\theta/2).\label{los}
 \ee The action of such an arbitrary beam splitter on an arbitrary joint input state $\ket{j,a}$ in Schwinger notation, is simply given as~\cite{noh},
 \be
 e^{i\theta\hat{J}_x}\ket{j,a}=\sum_{b=-j}^j e^{i\theta(b-a)}d_{a,b}^j(\theta)\ket{j,b},
 \label{formu}
 \ee
 where $d_{a,b}^j(\theta)$ is the usual rotational matrix element given as: 
 \ba
  d^{j}_{a,b}(\theta)=(-1)^{a-b}2^{-a}\sqrt{\frac{(j-a)!(j+a)!}{j-b)!(j+b)!}}P^{(a-b,a+b)}_{j-a}(\cos\theta)\nonumber \\ \times (1-\cos\theta)^{\frac{a-b}{2}}(1+\cos\theta)^{\frac{a+b}{2}},\;\;\;\;\;\;\;\;\;\;\;\;\;\;\;\;\;\;\;\;\;\;\;\;\;\;\;\ & &
  \label{rotmat}
  \ea
   where $P_n^{(\alpha,\beta)}(x)$ is the Jacobi polynomial~\cite{ang}. Also it is worth noting that when converted to Fock-basis, a two mode joint state in Schwinger notation, $\ket{j,a}$ is $\ket{j+a}\ket{j-a}$.
 
 The optimal state was originally derived by Berry and Wiseman~\cite{berry} conditioned on minimizing the phase variance with the canonical probability distribution given in Eq.~(\ref{prob}). Formally the optimal state is,
 \be
 \ket{\psi}_{\text{opt}}=\sum_{\mu=-j}^j\frac{1}{\sqrt{j+1}}\sin\left[\frac{(\mu+j+1)\pi}{2j+2}\right]\ket{j\mu}_y.
 \ee
Recall that the simultaneous eigenstate of $\hat{J}^2$ and $\hat{J}_y$ denote the state {\it within} the interferometer,  and thus the above state is after the first beam splitter, while $\ket{j+\mu}_{a}$ and $\ket{j-\mu}_{b}$ represents the fock state corresponding to modes $\hat{a}$ and $\hat{b}$ respectively. Rewriting the above state in a more explicit form as the product of the states at the two arms of the interferometer as,
 \be
 \ket{\psi}_{\text{opt}}=\sum_{\mu=-j}^j\psi_{\mu}\ket{j+\mu}_a\ket{j-\mu}_b,
 \ee
  where
  \be
   \psi_{\mu}=\frac{1}{\sqrt{j+1}}\sin\left[\frac{(\mu+j+1)\pi}{2j+2}\right]. \nonumber
 \ee
With the loss in mode $\hat{a}$, which is represented by the fictitious beam splitter, and $\ket{0}_c$ is the state entering the other input port, the combined input state for the fictitious beam splitter is: $\ket{j+\mu}_a\ket{0}_c$, and thus the state to be considered is,
\be
\ket{\psi}=\ket{\psi}_{\rm opt}\otimes\ket{0}_c=\sum_{\mu=-j}^j\psi_{\mu}\ket{j+\mu}_a\ket{j-\mu}_b\ket{0}_c.
\ee
The fictitious beam-splitter transforms modes $\hat{a}$ and $\hat{c}$ to modes $\hat{a}'$ and $\hat{c}'$ respectively.
Making use of the Schwinger representation for modes $\hat{a}$ and $\hat{c}$, the input $\ket{j+\mu}_a\ket{0}_c$ for the fictitious beam splitter can be written as a joint state: $\ket{\frac{j+\mu}{2},\frac{j+\mu}{2}}_{a,c}$. Letting $k=(j+\mu)/2$ and using Eq.~(\ref{formu}) we have the output as,
\ba
e^{i\theta\hat{J}_x}\ket{k,k}_{a,c}=\sum_{m=-k}^ke^{i\frac{\pi}{2}(m-k)}d_{mk}^k(\theta)\ket{k,m}_{a'c'} \;\;\;\;\ \nonumber \\ =\sum_{m=-k}^ke^{i\frac{\pi}{2}(m-k)}d_{mk}^k(\theta)\ket{k+m}_{a'}\ket{k-m}_{c'}.\;\ 
\ea
 Therefore the pure state of the inner modes $\hat{a}'$, $\hat{b}$ of the interferometer, and mode $\hat{c}'$ of the lost photons is given as,
\ba
 \ket{\psi}=  \sum_{\mu=-j}^j\sum_{m=-k}^k\psi_{\mu}e^{i\frac{\pi}{2}(m-k)}d_{mk}^k(\theta)\;\;\;\;\;\;\;\;\;\;\;\;\;\;\;\;\ \nonumber
 \\ \times \ket{k+m}_{a'}\ket{k-m}_{c'}\ket{j-\mu}_b,
 \label{pure}
 \ea 
where $k=(j+\mu)/2$ and $d_{mk}^k(\theta)$ is the usual rotational matrix element, as defined earlier.
\begin{widetext}
\section{ Density matrix description}
\label{densit}
The state specified in Eq.~(\ref{pure}) is a pure state and cannot be used further, so we need to calculate the reduced density matrix, by tracing out, mode $\hat{c}'$, as we have no more access to the lost photons. Thus first we need to calculate the total density matrix, representing the pure state of Eq.~(\ref{pure}),

\begin{eqnarray}
 \rho=\ket{\psi}\bra{\psi} 
  \;\;\;\;\;\;\;\;\;\;\;\;\;\;\;\;\;\;\;\;\;\;\;\;\;\;\;\;\;\;\;\;\;\;\;\;\;\;\;\;\;\;\;\;\;\;\;\;\;\;\;\;\;\;\;\;\;\;\;\;\;\;\;\;\;\;\;\;\;\;\;\;\;\;\;\;\;\;\;\;\;\;\;\;\;\;\;\;\;\;\;\;\;\;\;\;\;\;\;\;\;\;\;\;\;\;\;\;\;\;\;\;\;\;\;\;\;\;\;\;\;\;\;\;\;\;\;\;\;\;\;\;\;\;\;\;\;\;\;\;\;\;\;\;\;\;\;\;\;\;\;\;\;\;\;\;\;\;\;\;\;\;\;\;\;\;\;\;\;\;\;\;\ \nonumber \\
 =\sum_{\mu,\nu=-j}^j\sum_{m=-k}^k\sum_{n=-k'}^{k'}\psi_{\mu}\psi_{\nu}d_{m,k}^k(\theta)d_{n,k'}^{k'}(\theta)  e^{i\frac{\pi}{2}(m-k)}e^{-i\frac{\pi}{2}(n-k')}
 \ket{k+m}_{a'}\ket{k-m}_{c'}\ket{j-\mu}_{b{\hspace{1 mm}}a'}\bra{k'+n}_{c'}\bra{k'-n}_b\bra{j-\nu},\;\;\;\;\;\;\;\  
 \label{den}
 \end{eqnarray}

 where $k'=(j+\nu)/2$. The total density matrix given in Eq.~(\ref{den}) explicitly represents the state within the interferometer for a given loss, characterized by the angle $\theta$, and is useful in analyzing lossy interferometers such as the present one.
 
 As the mode $\hat{c}'$ is to be ignored, we need the reduced density matrix by tracing out that mode from the total density matrix given in Eq.~(\ref{den}). Thus we have,
 \begin{eqnarray}
 \rho'=\text{Tr}_{c'}[\ket{\psi}\bra{\psi}] 
  \;\;\;\;\;\;\;\;\;\;\;\;\;\;\;\;\;\;\;\;\;\;\;\;\;\;\;\;\;\;\;\;\;\;\;\;\;\;\;\;\;\;\;\;\;\;\;\;\;\;\;\;\;\;\;\;\;\;\;\;\;\;\;\;\;\;\;\;\;\;\;\;\;\;\;\;\;\;\;\;\;\;\;\;\;\;\;\;\;\;\;\;\;\;\;\;\;\;\;\;\;\;\;\;\;\;\;\;\;\;\;\;\;\;\;\;\;\;\;\;\;\;\;\;\;\;\;\;\;\;\;\;\;\;\;\;\;\;\;\;\;\;\;\;\;\;\;\;\;\;\;\;\;\;\;\;\;\;\;\;\;\;\;\;\;\;\ \nonumber \\
 =\sum_{\mu,\nu=-j}^j\sum_{m=-k}^k\sum_{n=-k'}^{k'} \psi_{\mu}\psi_{\nu}d_{m,k}^k(\theta)d_{n,k'}^{k'}(\theta))e^{i\frac{\pi}{2}(m-k)}e^{-i\frac{\pi}{2}(n-k')}\ket{k+m}_{a'}\ket{j-\mu}_{b\hspace{1 mm}a'}\bra{k'-n}_b\bra{j-\nu}\left [_{c'}\la{k-m}|{k'-n}\ra_{c'}\right].\;\;\;\;\;\;\
 \label{rden}
 \end{eqnarray}
 %Figure giving the result 
\begin{figure*}[tp]
\includegraphics[width=1.0\hsize,height=0.25\hsize]{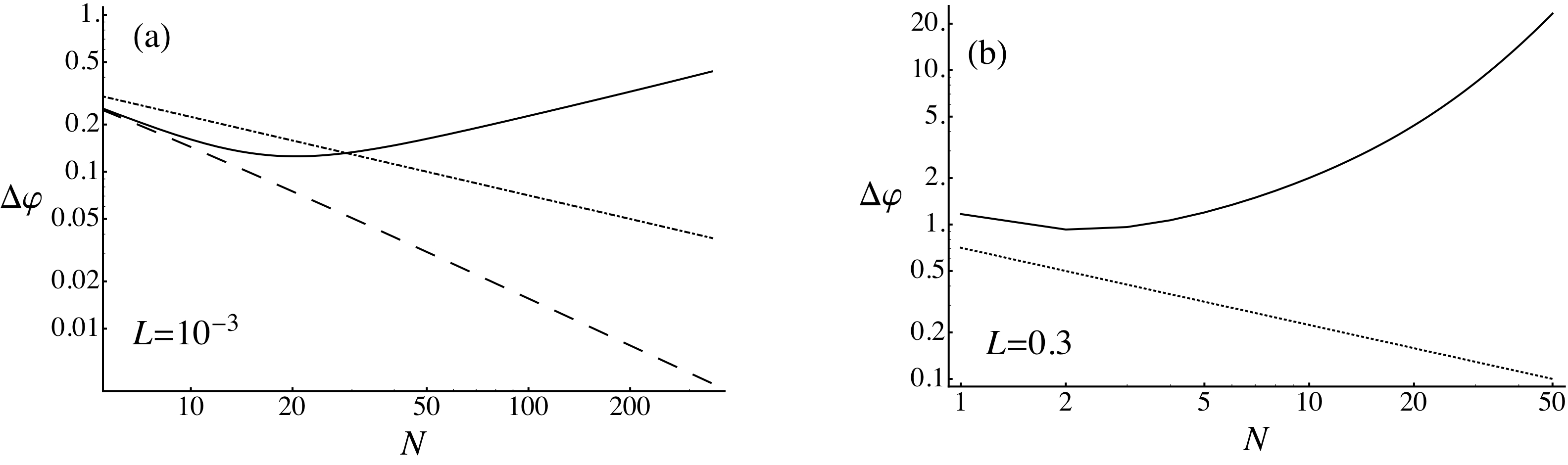}
\caption{Log-Log graph of minimum detectable phase versus the input photon number $N=2j$ for two different values of loss. (a) $L=10^{-3}$ and (b) $L= 0.3$. The dotted line in both figures is the shot noise limit $1/\sqrt{2j}$ and the dashed line in (a) is the Heisenberg limit given be Eq.~(\ref{hvar}), for comparison.The solid lines are numerically evaluated using Eq.~(\ref{varia}) and Eq.~(\ref{sharp}). Depending on the loss, the minimum detectable phase starts to diverge at certain photon numbers, which we shall call the {\it optimal number}, $N_{\rm opt}$.}
\label{final}
\end{figure*}

 Noting that $_{c'}\la{k-m}|{k'-n}\ra_{c'}=\delta_{k-m,k'-n}$, which eliminates the two exponential terms in Eq.~(\ref{rden}). This leads to:
 \ba
 \rho'=\sum_{\mu,\nu=-j}^j\sum_{m=-k}^k\sum_{n=-k'}^{k'} \psi_{\mu}\psi_{\nu}d_{m,k}^k(\theta)d_{n,k'}^{k'}(\theta)\delta_{k-m,k'-n}\ket{k+m}_{a'}\ket{j-\mu}_{b\hspace{1 mm}a'}\bra{k'-n}_b\bra{j-\nu}.\;\;\;\;\;\;\;\;\;\;\;\;\;\;\;\;\;\;\;\;\;\;\;\;\;\;\;\;\;\;\;\;\;\;\;\;\;\;\;\;\;\;\;\;\;\;\;\;\
 \label{rdenf}
 \ea 
 We can now use Eq.~(\ref{rdenf}) in Eq.~(\ref{prob}) to obtain the probability distribution and thus the minimum detectable phase as a function of $\theta$-which characterizes the photon loss-and the input photon number $2j$.

 \end{widetext}

\section{Phase estimate in presence of photon loss}
\label{conc}
To get an estimate using Holevo phase variance, we need to calculate the probability distribution, $P(\vp)$ using Eq.(\ref{prob}). 
We thus, in presence of photon loss have,
\be
 P(\vp)=\text{Tr}[\rho'\hat{F}(\vp)]. 
 \label{nprob}
 \ee
 Note that $\vp$ is an estimate of the system phase $\phi$. 

The POVM, $\hat{F}(\vp)d\vp$ [Eq.~(\ref{opo})] is given in terms of the $\hat{J}_y$ eigen states. Noting that the $\hat{J}_y$ eigen states are the states of the modes {\it within} the interferometer, we can rewrite the POVM as,
\be
\hat{F}(\vp)=\frac{1}{2\pi}\sum_{\ap,\bt=-j}^j e^{i(\ap-\bt)\vp}\ket{j,\ap}_{y}\bra{j,\bt}.
\label{eopo1}
\ee
Since the joint state $\ket{j,\ap}_y$ is written as  $\ket{j+\ap}_{a'}\ket{j-\bt}_b$, we have,
\ba
\hat{F}(\vp)=\frac{1}{2\pi}\sum_{\ap,\bt=-j}^j e^{i(\ap-\bt)\vp}\ket{j+\ap}_{a'}\ket{j-\ap}_{b} \nonumber \\ \otimes _{a'}\bra{j+\bt}_b\bra{j-\bt}
\label{eopo}
\ea

%\begin{widetext}
After the fictitious beamsplitter which represents photon loss associated with propagation, the inner modes of the MZI are $\hat{a}'$ and $\hat{b}$, and so the $\hat{J}_y$ eigen states are the product states of these modes. Thus using Eq.~(\ref{eopo}) in Eq.~(\ref{nprob}) along with Eq.~(\ref{rdenf}) for the reduced density matrix, after carrying out the trace operation for the modes $\hat{a}'$ and $\hat{b}$, we have the probability distribution in presence of loss as,
\ba
P(\vp) =\frac{1}{2\pi}\sum_{\substack{\mu,\nu=-j \\ \ap,\bt=-j }}^j\sum_{m=-k}^k\sum_{n=-k'}^{k'} [\psi_{\mu}\psi_{\nu}d_{m,k}^k(\theta)d_{n,k'}^{k'}(\theta) \nonumber  \\ \times e^{i(\ap-\beta)\vp} {\la{k'-n|j+\ap}\ra}{\la{j+\bt|k+m}\ra}\nonumber \\ \times {\la{j-\bt|j-\mu}\ra}{\la{j-\nu|j-\ap}\ra}]
\ea
 
 Recalling $k'=(j+\nu)/2$,  $k=(j+\mu)/2$, and because $\bt =\mu$ and $\ap=\nu$ as a consequence of the last two inner products in the above equation, we get from the first two inner products $n=k'$ and $m=k$ respectively. Hence the probability distribution explicitly given as,

 \be
 P(\vp)=\frac{1}{2\pi}\sum_{\mu,\nu=-j}^j\psi_{\mu}\psi_{\nu}e^{i(\nu-\mu)\vp}d_{k,k}^k(\theta)d_{k',k'}^{k'}(\theta).
 \label{prf}
 \ee
With this probability distribution we can calculate any moments of the phase estimate as a function of loss, $\theta$. Here we  calculate the Holevo phase variance rather than the standard variance, because the optimal state is derived by minimizing the Holevo phase variance~\cite{berry}. Using Eq.~(\ref{prf}) we now calculate sharpness $|\la e^{i\vp} \ra|$ is given as,
\ba
|\la e^{i\vp} \ra|=\int_0^{2\pi}P(\vp)e^{i\vp}d\vp \nonumber \;\;\;\;\;\;\;\;\;\;\;\;\;\;\;\;\;\;\;\;\;\;\;\;\;\;\;\;\;\;\;\;\;\;\;\;\;\;\;\;\;\;\
\\
=\sum_{\mu,\nu=-j}^j\delta_{\nu,\mu-1}\psi_{\mu}\psi_{\nu}d_{k,k}^k(\theta)d_{k',k'}^{k'}(\theta), \;\;\;\;\;\;\;\;\;\;\;\;\;\;\
\label{sharp1}
\ea
where $\delta_{\nu,\mu-1}=\frac{1}{2\pi}\int_0^{2\pi}e^{(\nu-\mu+1)\vp}d\vp$. Also using the rotational matrix element in Eq.~(\ref{rotmat}), we have $d_{k,k}^k(\theta)d_{k',k'}^{k'}(\theta)=\left[\cos^2\frac{\theta}{2}\right]^{(k+k')}$ and invoking the definition of $k$ and $k'$ in Eq.~(\ref{sharp1}), we obtain

\be
|\la e^{i\vp} \ra|=\sum_{\mu=-j}^j\psi_{\mu}\psi_{\mu-1}\left[\cos^2(\frac{\theta}{2})\right]^{j+\mu-\frac{1}{2}},
\label{sharp}
\ee
\begin{figure}[ht]
\includegraphics[height=0.5\hsize]{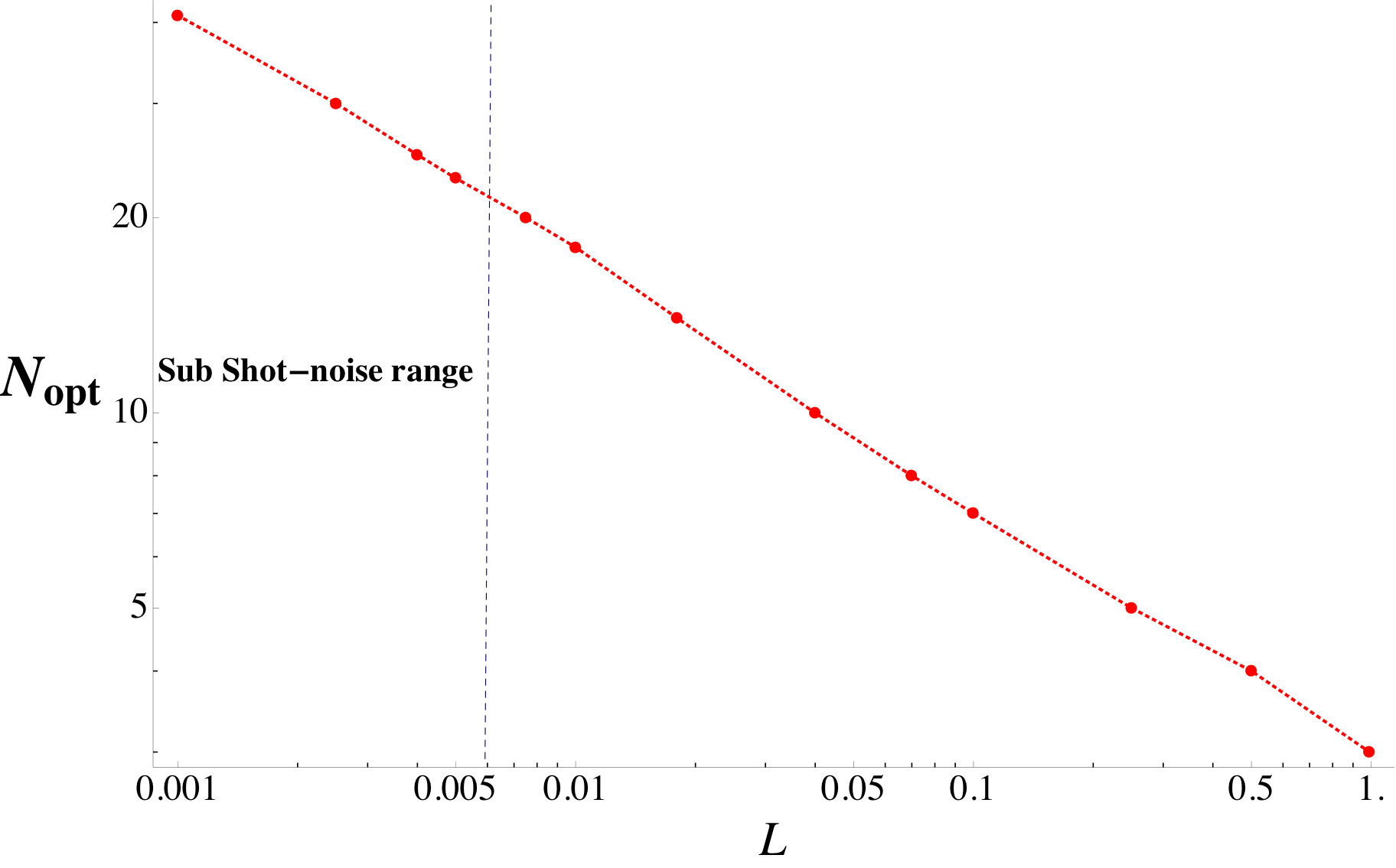}
\caption{Log-Log plot of the optimal number, $N_{\text{opt}}$ as a function of loss for various values of loss $L$. The sub-shot noise range indicates the range of loss, so as to obtain the sub-shot noise phase estimate}
\label{nl}
\end{figure} 
for the expression of the so-called sharpness as a function of loss. We use Eq.~(\ref{sharp}) in Eq.~(\ref{varia}) to obtain the phase estimate as a function of loss. 
\section{Numerical Results and Discussion}
\label{conc1}
Now the minimum detectable phase shift can be found with the uncertainty in phase estimate by plugging Eq.~(\ref{sharp}) in Eq.~(\ref{varia}). In Fig.~\ref{final} we numerically plot the  minimum detectable phase $\Delta\vp$ as a function of $2j=N$, the number of input photons, for two different values of loss, $L=10^{-3}$ and $L=0.3$. For comparison we also plot the Heisenberg limit for the optimal state given in Eq.~(\ref{hvar}). We can immediately observe that the minimum detectable phase is not always a continuously decreasing function, depending on the loss, the minimum detectable phase starts to diverge at certain photon numbers, which we shall call the {\it optimal number}, $N_{\text{opt}}$. 

As we do not have a closed form expression for the Holevo variance, as a function of loss $L$ and input photon number $N$, we have not found an analytical form for the optimal number, $N_{\text{opt}}$ as function of loss, $L$ [see Eq.(\ref{los})] . So we numerically plot $N_{\text{opt}}$ as a function of $L$ in Fig.~\ref{nl}  and it clearly decreases as the loss $L$ increases. 

Further, our analysis shows for small loss, about 0.05\% loss, which is equivalent to 99.5\% of transmission, we can still have the input number of photons, which gives a sub-shot noise limited phase estimate.

On the other hand, the upper bound for a certain number of input photons can be found where $\Delta\vp$ meets the shot-noise limit in Fig.~\ref{final}(a). That is to say, for a small amount of loss, we can still operate the interferometer, with the input number of photons less than the upper bound and achieve a  sub-shot noise limited phase estimate. 

To summarize, we analyzed the canonical phase measurement in presence of photon loss. Our formalism is based on the density matrix, which describes the mixed states, which naturally arise due to the presence of loss. Our analysis shows that the minimum detectable phase is not monotonically  decreasing function and would tend to increase at certain photon numbers, depending the loss present. Nevertheless, we can for small loss have considerably high number of input photons for the optimal state, that achieve sub-shot noise level phase estimates. We also have numerically plotted the optimal number, where the minimum detectable phase shift starts to diverge, as a function of loss, $L$, which would determine the optimal number of the input photon number for the optimal state, in presence of given amount of loss.

\section*{ Acknowledgments}We acknowledge financial support from the US Defense Advanced Research Projects Agency, the US Intelligence Advanced Research Projects Activity, and the US Army Research Office. We would like to acknowledge J. P. Dowling for helpful discussions.

\end{document}